\documentclass[twocolumn,prl,floatfix,superscriptaddress]{revtex4-1}

\usepackage[linktocpage,bookmarksopen,bookmarksnumbered]{hyperref}
\usepackage{graphicx} 
\usepackage{gensymb}
\usepackage{amsmath,bm,amssymb}
\usepackage{amsfonts}
\usepackage[usenames]{color}

\begin{document}

\title{Hund physics landscape of two-orbital system}
 \author{Siheon Ryee}
 \affiliation{Department of Physics, KAIST, Daejeon 34141, Republic of Korea}
 
 \author{Myung Joon Han}
 \email{mj.han@kaist.ac.kr}
 \affiliation{Department of Physics, KAIST, Daejeon 34141, Republic of Korea}
 
 \author{Sangkook Choi}
 \email{sachoi@bnl.gov}
 \affiliation{Condensed Matter Physics and Materials Science Department, Brookhaven National Laboratory, Upton, NY 11973, USA}
\date{\today}

\begin{abstract}
Motivated by the recent discovery of superconductivity in infinite-layer nickelates RE$_{1-\delta}$Sr$_\delta$NiO$_2$ (RE$=$Nd, Pr), we study the role of Hund's coupling $J$ in a quarter-filled two-orbital Hubbard model which has been on the periphery of the attention. A region of negative effective Coulomb interaction of this model is revealed to be differentiated from three- and five-orbital models in their typical Hund's metal active fillings. We identify distinctive regimes including four different correlated metals, one of which stems from the proximity to a Mott insulator while the other three, which we call ``intermediate" metal, weak Hund's metal, and valence-skipping metal, from the effect of $J$ being away from Mottness. Defining criteria characterizing these metals are suggested, establishing the existence of Hund's metallicity in two-orbital systems.

\end{abstract}
 
\maketitle

A novel route to the electron correlation, which has attracted a great deal of attention over the last fifteen years, is on-site Hund's coupling $J$ \cite{Georges}. This energy scale favors high-spin configurations on each atom lifting the degeneracy of atomic multiplets. In multiorbital systems away from half-filling, an intriguing correlated metallic regime dubbed Hund's metal emerges, promoted by $J$ rather than the proximity to a Mott insulator \cite{Haule,Yin1,Janus,Georges}. Accordingly, many related physical phenomena have been being uncovered such as the spin-freezing crossover \cite{Werner1,Hoshino}, the spin-orbital separation \cite{Stadler1,SOS1,SOS2,SOS3,SOS4,Deng,Stadler2}, instability to the charge disproportionation \cite{Isidori,Ryee}, the orbital differentiation \cite{OSMP1,Medici_2011,Bascones,Lanata,Medici_2014,Kostin,Kugler1}, and superconductivity \cite{Hoshino,THLee}, to name a few. These concepts have provided a compelling view of the physics, most prominently of iron-based superconductors \cite{Haule,Hansmann,Yin1,Yin2,Bascones,Lanata,nematicity,Medici_2017,Arribi,THLee,Belozerov} and ruthenates \cite{Mravlje,SRO,Kugler2,HJLee}.

In the midst of unveiling Hund's metal phenomenology, however, two-orbital models with one electron away from half-filling have been on the periphery of the attention, although intriguing effects of Hund's coupling have been reported \cite{Medici_2011,Janus,Stadler_thesis}. 
This is presumably because this usual filling for Hund's metallicity results in the seemingly trivial singly occupied electron/hole state for this case.

The recent discovery of the superconductivity in infinite-layer nickelates RE$_{1-\delta}$Sr$_\delta$NiO$_2$ (RE~$=$~Nd, Pr) \cite{Li,Osada} heralds a new chapter of quantum materials research \cite{KWLee,Botana,Nomura,Sakakibara,Jiang,XWu,Hepting,Goodge,Ryee2,Peiheng,Si,Werner_nickelate,Hu,Vishwanath,Adhikary,GMZhang,Gao,Bernardini,MYChoi,Karp,Lechermann2,Olevano,Kitatani,Dagotto,QGu,XWu2,HuZhang,Petocchi,Lechermann,YWang,CJKang,BKang,Wan,Rossi}. Despite their chemical and structural similarities with cuprates (nominal one hole occupation of Ni-$d$ orbitals residing in the NiO$_2$ plane), they exhibit sharp differences in their normal state physical properties. Most strikingly, they are metals without long-range magnetic orders showing non-Fermi-liquid behaviors at elevated temperatures \cite{Li,Li2,Osada,Zeng}. A series of the recent {\it ab initio} studies reported the importance of Hund's coupling \cite{Lechermann,YWang,CJKang,BKang,Wan}, especially among Ni-$e_g$ (two-orbital) electrons \cite{BKang}, alluding to an intriguing route to the superconductivity \cite{Hu,Vishwanath,Adhikary}. Although these observations are interesting per se, a suitable reference picture of Hund's physics has yet to be established.

To that end, in this work we classify distinctive regimes emerging out of two-orbital Hubbard model away from half-filling. Four different correlated metals are identified: one of which stems from the proximity to a Mott insulator while the other three from effects of $J$ being away from Mottness.
The latter three $J$-induced metals are ``intermediate", weak Hund's (WH), and valence-skipping (VS) metals. Characteristic features of these metals will be discussed throughout the paper. 
We finally discuss implications of our two-orbital picture to the physics of infinite-layer nickelates.
 
To obtain a basic picture, we first begin with a brief excursion into a simple atomic limit: a collection of atoms with zero hopping among them. We consider three different models: two-, three-, and five-orbital models with $n_d=M+1$ electron filling ($M$: the number of orbital). This particular choice is motivated by the observation that one electron away from the half-filling host Hund's metallicity when they form solids (at the least for $M\geq3$) as well as that each model is relevant to nickelates ($M=2$), ruthenates ($M=3$), and iron-based superconductors ($M=5$). We take the following form for the local Hamiltonian of $M=2$ and $3$ models:
\begin{align}
\begin{split}
H_\mathrm{loc} &=  U\sum_{m}{n_{m \uparrow} n_{m \downarrow}} 
+ \sum_{mm',\sigma\sigma'}^{m < m'}(U'-J\delta_{\sigma\sigma'}){ n_{ m \sigma} n_{ m' \sigma'}} \\
&+ J\sum_{mm'}^{m \neq m'}(d^{\dagger}_{m \uparrow}  d^{\dagger}_{m' \downarrow} d_{m \downarrow} d_{m' \uparrow} 
+d^{\dagger}_{ m \uparrow} d^{\dagger}_{ m \downarrow} d_{ m' \downarrow} d_{ m' \uparrow}) \\
& -\mu\sum_{m,\sigma}n_{m\sigma},
\end{split}
\label{eq1}
\end{align}
where $d^{\dagger}_{m \sigma}$ ($d_{m \sigma}$) is the electron creation (annihilation) operator with orbital index $m=1,...,M$ and spin index $\sigma=\uparrow, \downarrow$. $n_{m\sigma} = d^{\dagger}_{m \sigma}d_{m \sigma}$ is the number operator. $\mu$ is the chemical potential to obey average electron filling of $n_d=M+1$ per site. $U$ ($U'$) is intraorbital (interorbital) Coulomb energy cost. We set $U'=U-2J$ assuming cubic symmetry. For the $M=5$ case, the above Kanamori-type two-body terms are far from reality, and thus a well-suited strategy, e.g., Slater parametrization, is required. One possible way is to introduce the relative strength of anisotropic interaction ($1/\gamma$ where $\gamma>0$) and reparametrize Slater integrals in terms of it \cite{Strand}. In this way, $1/\gamma=0$ limit corresponds to Eq.~(\ref{eq1}) even for $M=5$ (see Supplemental Material (SM) \cite{supple}).

\begin{figure} [!htbp] 
	\includegraphics[width=0.9\columnwidth, angle=0]{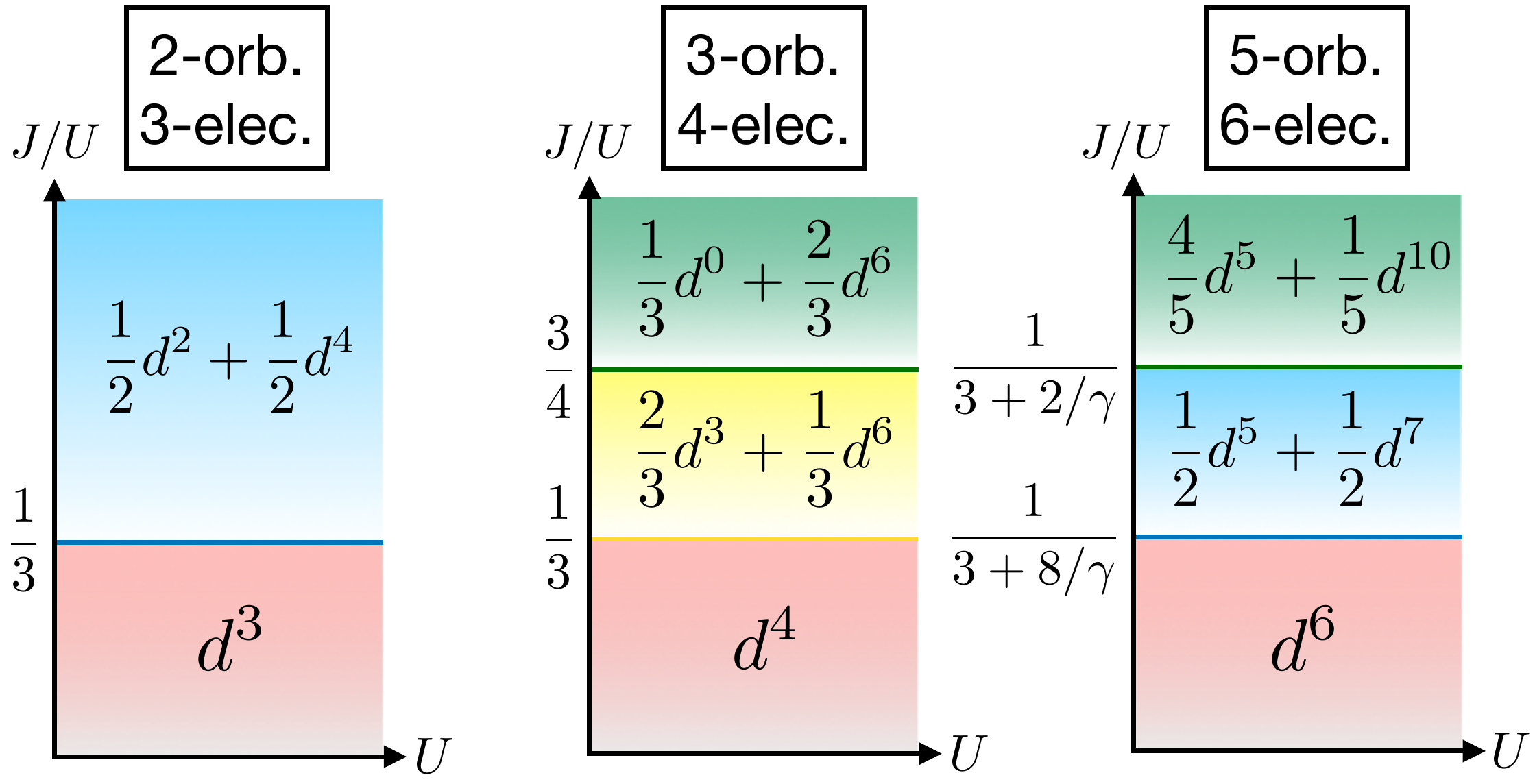}
	\caption{Atomic limit phase diagrams at $T=0$, $U>0$, and $J/U \geq 0$ for three different models ($M=2,3$, and $5$ systems) with $n_d=M+1$. The lowest-energy configurations are indicated at each region. The VS phase is highlighted with skyblue region. }
	\label{fig1}
\end{figure}

The ground state configurations of atomic limit at vanishing temperature ($T=0$) are presented in Fig.~\ref{fig1}. Here we use notation $cd^n$ to denote the ratio ($c$) of sites having $n$-electron occupation in the configuration. The homogeneous phases ($d^{n_d}$) occupy the small $J/U$ regions relevant to most of real materials. 

For large $J/U$, on the other hand, the mixed valence phases emerge. For all the cases with Kanamori interaction ($1/\gamma=0$ for $M=5$), the transition occurs from a homogeneous to a mixed valence state when $J/U>1/3$, i.e., $U-3J<0$.
Only the $M=2$ case shows VS transition ($d^N \rightarrow (d^{(N+1)} + d^{(N-1)})/2$) under this form of interaction. This VS phenomenon is the direct manifestation of the negative $U_\mathrm{eff}$: $U_\mathrm{eff}\equiv E_{N+1}+E_{N-1}-2E_{N}<0$ where $E_N$ is energy of the lowest-lying $N$-electron state \cite{Anderson,Katayama,Varma}. 
The $1/\gamma \neq 0$ case of $M=5$ also leads to VS, albeit an extreme form of mixed valence preempts the region of $J/U > 1/(3+2/\gamma)$ masking the VS phase (see the rightmost panel in Fig.~\ref{fig1} or see Ref.~\cite{Strand} for the $M=5$ case). 
To summarize, we identify $M=2$ case as the {\it minimal} model for $J$-driven VS phenomenon.

With insight obtained above, we now turn to the actual lattice problem with nonzero hopping. In order to focus on generic features rather than material specific ones, we consider the infinite dimensional Bethe lattice of semicircular density of states with half-bandwidth $D=1$. $D$ is hereafter used as the unit of energy. We solve the $M=2$ case with $n_d=3$ (particle-hole symmetric about $N=2$). The interaction form of Eq.~(\ref{eq1}) is used for non-hybridized degenerate two orbitals. The model is solved within the dynamical mean-field theory (DMFT) \cite{DMFT} employing \textsc{comctqmc} implementation \cite{Choi} of the hybridization-expansion continuous-time quantum Monte Carlo \cite{CTQMC} as an impurity solver. Unless otherwise specified, $T=0.01$. We restrict ourselves to paramagnetic solutions without spatial symmetry-breaking.

Central physical quantity of the present study is the onset temperatures of screening of spin and orbital degrees of freedom. These two temperatures, $T^\mathrm{onset}_\mathrm{spin}$ and $T^\mathrm{onset}_\mathrm{orb}$, are defined as the temperature below which the Curie law of unscreened local spin/orbital moment starts to become violated and screening sets in \cite{Deng}. A hallmark of strong Hundness is the separation of these two temperatures: $T^\mathrm{onset}_\mathrm{orb} \gg T^\mathrm{onset}_\mathrm{spin}$ yielding a range of temperature in which unscreened local spin moment coexists with the screened orbital degrees of freedom \cite{Okada,SOS1,SOS2,SOS3,SOS4,Stadler1,Stadler2,Deng}. We will measure the separation of two $T^\mathrm{onset}$ as $\Delta T^\mathrm{onset} \equiv T^\mathrm{onset}_\mathrm{orb} - T^\mathrm{onset}_\mathrm{spin}$. 
To locate the onset temperatures, we first evaluate the local spin/orbital susceptibilities: $\chi_\mathrm{s/o}=\int_{0}^{1/T}{d\tau \big( \langle {O}_\mathrm{s/o}(\tau){O}_\mathrm{s/o} \rangle - \langle {O}_\mathrm{s/o} \rangle^2 \big)}$, where ${O}_\mathrm{s}(\tau)=\sum_{m}n_{m\uparrow}(\tau)-n_{m\downarrow}(\tau)$ for spin and ${O}_\mathrm{o}(\tau)=\sum_{\sigma}n_{1\sigma}(\tau)-n_{2\sigma}(\tau)$  for orbital ($\tau$: imaginary time) up to $T=1$, and then fit high-$T$ data to the following formula: $\chi_\mathrm{s/o} \propto 1/(T+T^\mathrm{onset}_\mathrm{spin/orb})$.

\begin{figure} [!htbp] 
	\includegraphics[width=0.97\columnwidth, angle=0]{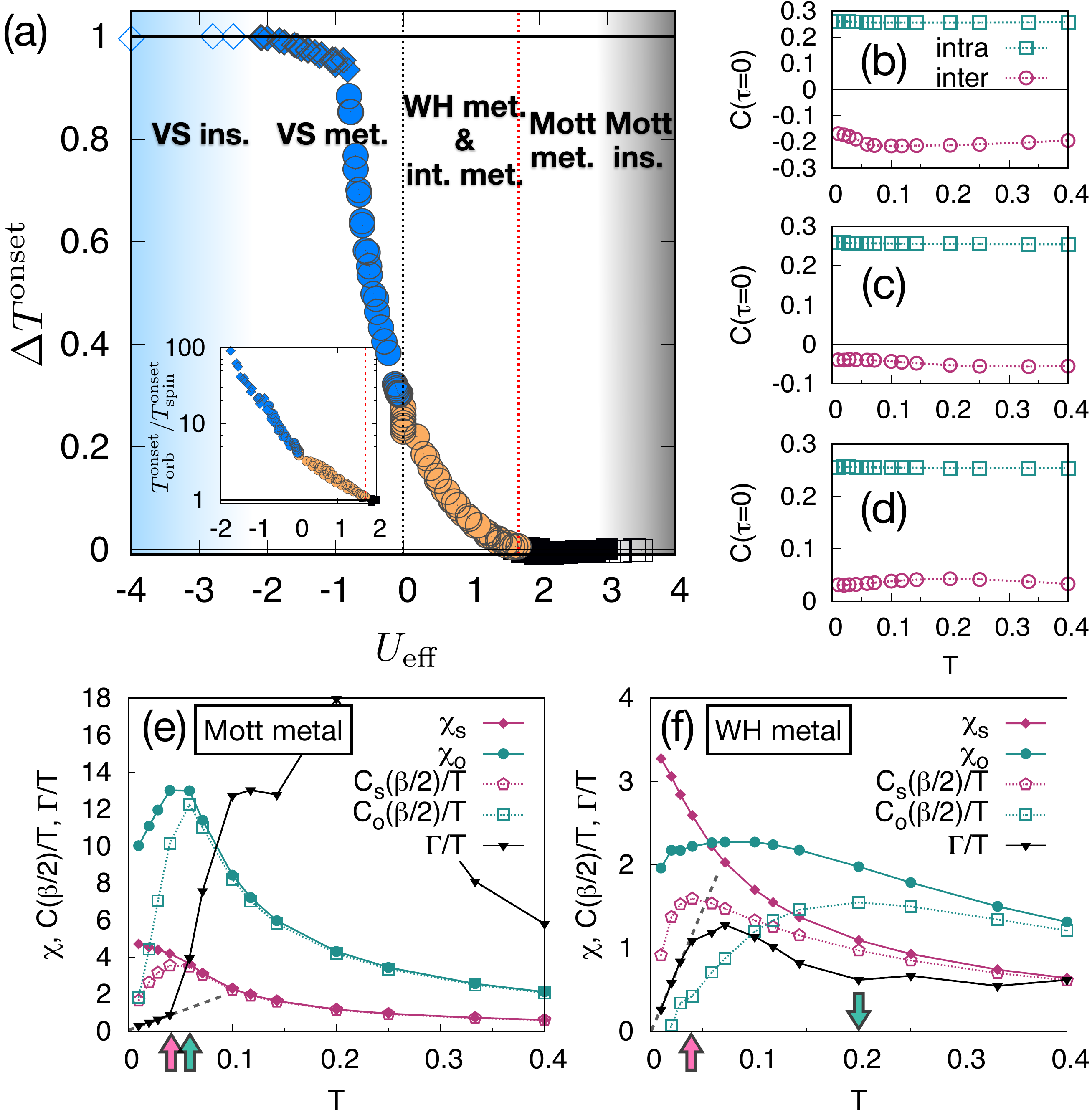}
	\caption{(a) $\Delta T^\mathrm{onset}$ as a function of $U_\mathrm{eff}$ obtained from $2\leq U/D \leq 8$ and $0 \leq J/U \leq 0.5$. The red dotted line indicates the approximate value of $U_\mathrm{eff}$ below which $\Delta T^\mathrm{onset}>0$. The black dotted line denotes $U_\mathrm{eff}=0$. We used $T=0.05$ to stabilize the VS insulator phase \cite{comment1}. The diamond symbols denote the region in which $T^\mathrm{onset}_\mathrm{orb}>1$. Inset: $T^\mathrm{onset}_\mathrm{orb}/T^\mathrm{onset}_\mathrm{spin}$ as a function of $U_\mathrm{eff}$.
	(b--d) $C_\mathrm{intra}$ (green square) and $C_\mathrm{inter}$ (magenta circle) at $U=4$ for (b) Mott metal ($J/U=0.15$; $U_\mathrm{eff}=2.2$), (c) WH metal ($J/U=0.3$; $U_\mathrm{eff}=0.4$), and (d) VS metal ($J/U=0.37$; $U_\mathrm{eff}=-0.44$).
	(e--f) $\chi_\mathrm{s/o}$, $\Gamma/T$, and $C_\mathrm{s/o}(\beta/2)/T$ plotted as a function of $T$ at $U=4$ for (e) Mott metal ($J/U=0.15$; $U_\mathrm{eff}=2.2$) and (f) WH metal ($J/U=0.3$; $U_\mathrm{eff}=0.4$). $T^\mathrm{peak}_\mathrm{spin/orb}$ are marked with magenta (spin) and green (orbital) arrows. The grey dashed lines are guide to the eye to indicate quasilinearity of $\Gamma/T$ (i.e., $\Gamma \sim T^2$). }
	\label{fig2}
\end{figure}

Figure~\ref{fig2}(a) presents our calculated $\Delta T^\mathrm{onset}$ as a function of $U_\mathrm{eff}$. 
Most interestingly, we found a {\it generic} scaling relation between $\Delta T^\mathrm{onset}$ and $U_\mathrm{eff}$. 
Note also that $T^\mathrm{onset}_\mathrm{orb}/T^\mathrm{onset}_\mathrm{spin}$ clearly shows the same trend as shown in the inset of Fig.~\ref{fig2}(a). This implies that $U_\mathrm{eff}$ ($U_\mathrm{eff}=U-3J$ for our case) is the crucial factor, rather than $U$ or $J$ alone, for the separation of two onset temperatures. This result is consistent with the recent comparative study of real materials \cite{Deng}, and demonstrates the generality holding for wide range of $U$ and $J/U$ in the two-orbital model.

By looking at Fig.~\ref{fig2}(a), one can first notice the presence of two distinctive types of insulators, namely the Mott and valence-skipping (VS) insulators at large positive and negative values of $U_\mathrm{eff}$, respectively. 
The former is the result of suppressed charge fluctuations localizing electron motions, thereby maximizing the probability of $|N=n_d=3,S=1/2\rangle$ multiplets ($N$: charge, $S$: spin). By contrast, the latter form of insulator exhibits the predominance of two multiplets, $|2,1\rangle$ and $|4,0\rangle$ with largely suppressed $|3,1/2\rangle$ probability because $U_\mathrm{eff}<0$ \cite{supple}. The presence of these two phases is reminiscent of the atomic limit result (see Fig.~\ref{fig1}).

Interestingly, we identify distinctive regimes within metallic phase intervening between the two insulators. 
When $U\gg J$, a metal with $\Delta T^\mathrm{onset} \simeq 0$ is found to appear near a Mott insulator where Mottness dominates over Hundness (see Fig.~\ref{fig2}(a)). 
To gain some understanding of this behavior, we resort to a low-energy Kondo model by performing a Schrieffer-Wolff transformation of relevant impurity Hamiltonian \cite{SWT}. The resulting Kondo coupling constants ($\mathcal{J}^i$) when $U>J$ read $\mathcal{J}^i_{U,J} \simeq \mathcal{J}^i_{U,J=0} + \mathcal{O}(\frac{J}{U^2}) + \mathcal{O}(\frac{J^2}{U^3}) + \cdots$ because $\mathcal{J}^i_{U,J} \sim V^2/{\Delta E}$. Here, $V$ is the bath-impurity hybridization strength and $\Delta E$ is the charge excitation energy from $N=n_d$ to $N=n_d\pm1$ subspaces. In the regime of $U \gg J$, $\mathcal{J}^i_{U,J} \simeq \mathcal{J}^i_{U,J=0}$ by which the system approximates to a $\mathrm{SU(4)}$ model having $\mathcal{J}^\mathrm{spin} = \mathcal{J}^\mathrm{orb}$. In this case, the relation $\mathcal{J}^\mathrm{spin} = \mathcal{J}^\mathrm{orb}$ also holds under renormalization group flow \cite{Kuramoto,SOS3}, thereby the Kondo screening of spin and orbital occur simultaneously. In this respect, we identify a regime of strong Mottness ($U\gg J$) with $\Delta T^\mathrm{onset} \simeq 0$. Following the terminology of Ref.~\cite{Deng,Stadler2}, we call the metallic regime of $\Delta T^\mathrm{onset} \simeq 0$ a Mott metal (see Fig.~\ref{fig2}(a)).

On the contrary, there exist metals with a finite $\Delta T^\mathrm{onset}$. Near VS insulator where $U_\mathrm{eff}<0$, a correlated metal emerges exhibiting a tendency of valence-skipping, which we call a VS metal (Fig.~\ref{fig2}(a)). 
In order to characterize this metal, we examine the sign of $C_\mathrm{inter}$ ($C_\mathrm{intra/inter}=\langle \delta n_m \delta n_{m'} \rangle$ where $\delta n_m = \sum_{\sigma}n_{m\sigma}- \langle \sum_{\sigma}n_{m\sigma} \rangle$ and $m\neq m'$ for $C_\mathrm{inter}$ while $m=m'$ for $C_\mathrm{intra}$). 
As VS metal emerges when $U_\mathrm{eff}<0$, multiplets in $N={2}$ and $N={4}$ charge subspaces are lower in energy than those of $N={3}$. Thus, either electrons or holes try to occupy both orbitals yielding $C_\mathrm{inter}>0$ in contrast to the case of metals belonging to $U_\mathrm{eff}>0$ (compare Fig.~\ref{fig2}(d) with (b) and (c)). 
Due to this negativity of $U_\mathrm{eff}$, VS region is highly susceptible to the formation of charge disproportionation \cite{Strand,Ryee} or superconductivity \cite{Micnas,Hoshino}, thereby being detectable when accompanied by such orders.

We now turn to the metallic region where $U_\mathrm{eff}>0$ with $\Delta T^\mathrm{onset}>0$ which is of our particular interest due to the potential presence of Hund's metallicity. Although strong Hundness is argued to be manifested by $\Delta T^\mathrm{onset}>0$ \cite{Deng}, the presence of Hund's metal regime in our two-orbital model has yet to be established. The emergence of Hund's metallicity has been attributed to the two-faced effect of $J$ dubbed ``Janus effect" suppressing the quasiparticle weight $Z$ \cite{comment2} on one hand, while enhancing $U_c$ ($U_c$: the critical value of $U$ for the Mott transition) on the other hand \cite{Janus}.
Thereby it reflects the stronger correlation (i.e, reduced $Z$) induced by $J$ rather than the proximity to a Mott insulator. This effect has been clearly seen in systems with $n_d=M\pm1$ among $M\ge3$ orbitals in which the atomic ground state degeneracy is lifted by $J$ \cite{Janus,Georges}.
On the contrary, our two-orbital system with $n_d=M+1=3$ hosts single hole, whereby it has been a conventional wisdom that the Janus effect is absent in two-orbital models away from half-filling.

Here we argue that, albeit weak, the Janus effect can be identified even in the two-orbital case. We first note that the conventional way of capturing this effect is to plot the evolution of $Z$ as a function of $U$, and then to examine whether the suppression of $Z$ and enhancement of $U_c$ simultaneously occur at a fixed $U$ as $J$ is increased, i.e., examining whether $(\partial{Z}/\partial{J})_{U}<0$ and $d{U_c(J)}/dJ>0$. In this strategy, however, the interorbital Coulomb energy cost, $U'=U-2J$, is not fixed as $J$ is varied. Thus, the genuine effect of tuning $J$ is partly masked by the reduced $U'$.

\begin{figure} [!htbp] 
	\includegraphics[width=1.0\columnwidth, angle=0]{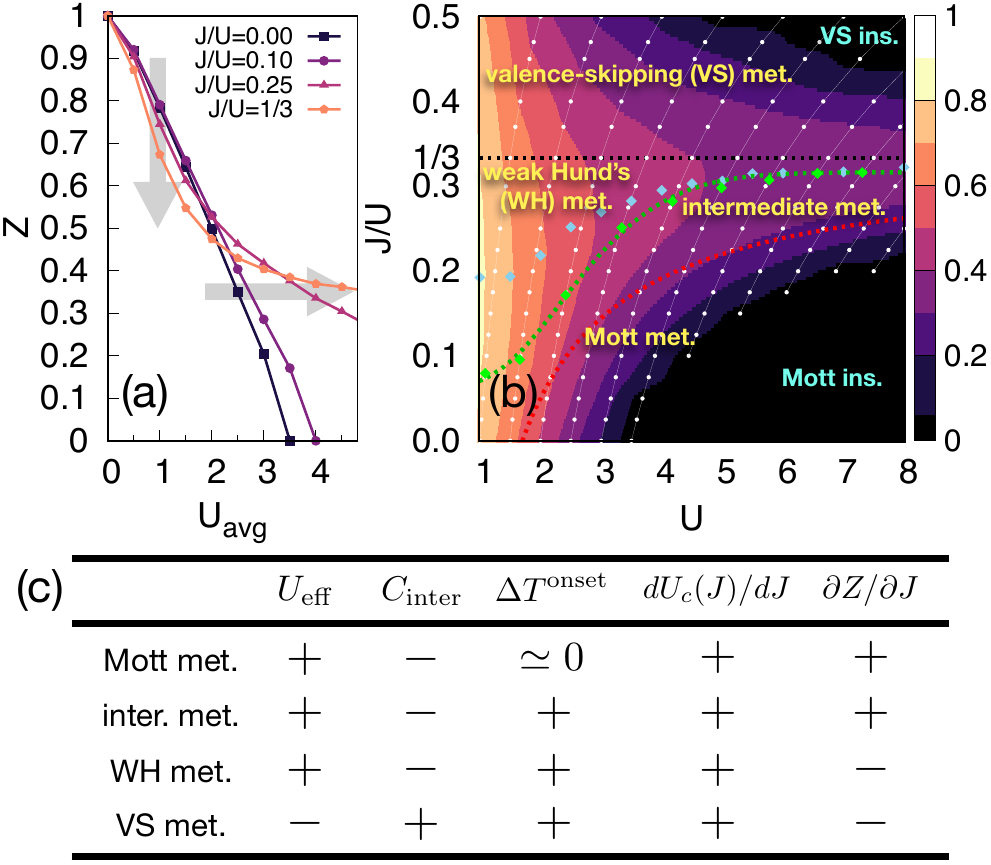}
	\caption{(a) $Z$ as a function of $U_\mathrm{avg}~(=U-J)$ for several $J/U~(=\frac{J/U_\mathrm{avg}}{1+J/U_\mathrm{avg}})$. (b) Phase diagram with color scheme representing $Z$. The red and black dotted lines indicate the same as in Fig.~\ref{fig2}(a).
	The green dotted line is an estimated boundary of $J/U$ above which the Janus effect emerges: $(\partial{Z}/\partial{J})_{U_\mathrm{avg}}<0$ while $d{U_c}(J)/d{J}>0$. Green diamonds are actual crossing points where $(\partial{Z}/\partial{J})_{U_\mathrm{avg}}=0$, whereas blue ones are $(\partial{Z}/\partial{J})_U=0$ plotted for comparison.
	A set of $U$ and $J/U$ values belonging to the same $U_\mathrm{avg}$ are connected with a white solid line. 
	(c) Characteristic features of different correlated metals. Here, $+/-$ denote the sign of the corresponding quantity. 
	}
	\label{fig3}
\end{figure}

To circumvent the above difficulty, we propose an alternative ``gauge" of measuring the Janus effect: examining the sign of $(\partial{Z}/\partial{J})_{U_\mathrm{avg}}$ where $U_\mathrm{avg} \equiv {1}/{M^2} \sum_{mm'}\mathcal{U}_{m m'}$ ($\mathcal{U}_{m m'}$: elements of Coulomb interaction tensor; see SM \cite{supple}). The rationale behind this proposal is that one should count not only the intra- ($m=m'$), but also the interorbital ($m\neq m'$) Coulomb energy cost which may vary with $J$. 
For the current $M=2$ case, $U_\mathrm{avg}=U-J$. It is worth noting that within the Slater parametrization for $M=5$, $U_\mathrm{avg}=F^0$ ($F^0$: the zeroth-order Slater integral which is the monopole term of Coulomb interaction). We also point out that this kind of viewpoint is implicitly embodied in some {\it ab initio} studies (e.g., Ref.~\cite{Haule,CJKang}) by the use of Slater parametrization of the Coulomb interaction with $F^0$ remaining unchanged while varying $J$ in searching for the Hund's physics. With this idea in mind, we plot $Z$ vs. $U_\mathrm{avg}$ in Fig.~\ref{fig3}(a). One can now clearly capture the Janus effect; namely, $(\partial{Z}/\partial{J})_{U_\mathrm{avg}}<0$ and $d{U_c}(J)/d{J}>0$. 
We suspect that this reduction of $Z$ by $J$ is attributed to the lifted degeneracy in $N=2$ charge subspace. In this subspace, $J$ elevates the energy of $|2,0\rangle$ states and lowers that of $|2,1\rangle$. The enhanced fluctuation between $|2,1\rangle$ and $|3,1/2\rangle$ suppresses $\mathcal{J}^\mathrm{spin}$ as in the case of $M\geq3$ models with $n_d=M\pm1$ \cite{Yin1,SOS1,SOS2,SOS3}.
In the meanwhile, atomic state of $n_d$ subspace ($N=3$) is not affected by $J$ since it plays no role when single electron/hole is occupied. As a result, overall influence of $J$ is weaker than the $M\geq3$ cases. In this sense, we call our metallic regime satisfying $(\partial{Z}/\partial{J})_{U_\mathrm{avg}}<0$ (or $(\partial{Z}/\partial{J})_{U}<0$) and $d{U_c}(J)/d{J}>0$ a ``weak" Hund's (WH) metal (Fig.~\ref{fig3}(b)).

As a central result of our study, we present the phase diagram exhibiting different metallic regimes; see Fig.~\ref{fig3}(b). In Fig.~\ref{fig3}(c), we also summarize characteristic features of these correlated metals. 
The green dotted line in Fig.~\ref{fig3}(b) denotes $J/U$ above which the Janus effect exists. Hence we now further classify the region of $U_\mathrm{eff}>0$ and $\Delta T^\mathrm{onset}>0$ into two: WH metal exhibiting the Janus effect and intermediate metal which emerges in an ``intermediate" region between the WH and Mott metals. In the intermediate metal, $J$ alleviates the correlation strength, i.e., $(\partial{Z}/\partial{J})_{U_\mathrm{avg}}>0$, although spin-orbital separation ($\Delta T^\mathrm{onset}>0$) occurs.

Notable feature of this phase diagram is that near $J/U=1/3$ line which is the boundary between WH and VS metals, quasiparticle survives up to an arbitrarily large $U$. Indeed, $(\partial Z/\partial U_\mathrm{avg})_{J/U=1/3} \rightarrow 0$ while $Z$ remains small but finite as $U_\mathrm{avg}$ is increased (Fig.~\ref{fig3}(a)). This is because the lowest-energy atomic multiplets in $N=n_d$ and $N=n_d\pm1$ subspaces are degenerate or sufficiently close in energy around this line resulting in $U_\mathrm{eff}=U-3J \simeq0$. As a result, hopping processes become feasible, which otherwise should be blocked due to a large Coulomb energy cost \cite{Isidori}.

Having established an overall picture, we now further examine the spin-orbital separation via
long-time spin/orbital correlators at $\tau=\beta/2$ ($\beta=1/T$): $C_\mathrm{s/o}(\beta/2)=\langle {O}_\mathrm{s/o}(\beta/2) {O}_\mathrm{s/o} \rangle - \langle {O}_\mathrm{s/o} \rangle^2$ in Fig.~\ref{fig2}(e--f). 
At sufficiently low temperatures, $C_\mathrm{s/o}(\beta/2)$ scales as $C_\mathrm{s/o}(\beta/2) \sim \big(T/(\mathrm{sin}(\pi/2)) \big)^\alpha$ with $\alpha=2$ in a Fermi-liquid, while $\alpha=1$ in the crossover between local moment and the Fermi-liquid \cite{Werner1,Cha}. In the unscreened local moment regime, $C_\mathrm{s/o}(\beta/2)$ is basically $T$-independent and $\chi_\mathrm{s/o} \simeq C_\mathrm{s/o}(\beta/2)/T$ (Fig.~\ref{fig2}(e--f)). In light of this observation, we investigate $C_\mathrm{s/o}(\beta/2)/T$ for an extended range of $T$. These quantities should be $T$-linear in the Fermi-liquid, whereas scale as $1/T$ in the local moment regime. As a consequence, a narrow region of crossover between these two emerges forming a hump of $C_\mathrm{s/o}(\beta/2)/T$. Temperatures at which peaks of $C_\mathrm{s/o}(\beta/2)/T$ appear ($T^\mathrm{peak}_\mathrm{spin/orb}$) are marked with arrows in Fig.~\ref{fig2}(e--f). One can clearly notice that the $T^\mathrm{peak}_\mathrm{spin}$ coincides with $T^\mathrm{peak}_\mathrm{orb}$ in the Mott metal whereas two crossover temeratures become separated in the metals with finite $\Delta T^\mathrm{onset}$ such as WH metal. Furthermore, below $T^\mathrm{peak}_\mathrm{spin}$ the quasiparticle scattering rate $\Gamma = -Z\mathrm{Im}[\Sigma(i\omega)]\big|_{\omega \rightarrow 0}$ roughly follows the Fermi-liquid behavior ($\Gamma \propto T^2$). This result is consistent with the observation that $C_\mathrm{s}(\beta/2)/T$ is sublinear in $T$ when $T<T^\mathrm{peak}_\mathrm{spin}$.

While we mainly focus on the generic features of Mott and Hund physics, its relevance to RE$_{1-\delta}$Sr$_\delta$NiO$_2$ is of particular interest. Our two orbitals can be regarded as Ni-$d_{x^2-y2}$ and another Ni-$d$ orbital. 
{\it Ab initio} estimate of Coulomb interaction for a Ni-$e_g$ model reads $U/D \simeq 1.7$ and $J/U \simeq 0.2$ \cite{Sakakibara}, which falls into the WH metal region (Fig.~\ref{fig3}(b)). However, $\Delta$, the on-site energy level splitting between two orbitals, and its competition with $J$ should also be taken into account. 
If $\Delta$ dominates over $J$, singlet $|2,0\rangle$ would be favored over triplet $|2,1\rangle$ hampering strong Hund's physics. In the presence of $\Delta$, eigenvalues of local Hamiltonian are $E_{|2,0\rangle} = U-\Delta-\sqrt{J^2+\Delta^2}-2\mu$ and $E_{|2,1\rangle}=U-3J-\Delta-2\mu$ for $|2,0\rangle$ and $|2,1\rangle$, respectively \cite{supple}. Hence, the criterion for predominance of $|2,1\rangle$ over $|2,0\rangle$ (i.e., $E_{|2,1\rangle}<E_{|2,0\rangle}$) is $J/\Delta>\sqrt{2}/4 \simeq 0.354$. 
However, estimated value $J/\Delta \simeq 0.3$ for Nd$_{0.8}$Sr$_{0.2}$NiO$_2$ between two Ni-$e_g$ \cite{Sakakibara,Lechermann} is slightly smaller than the ``bare" critical value $J/\Delta \simeq 0.354$. 
In this respect, nickelates may belong to the competing region where large $J/U$ favors Hund's metallicity while $J/\Delta$ which is slightly smaller than its threshold value refrains from forming high-spin in the two-hole atomic state (see SM for related DMFT results \cite{supple}). Thus, metallic nature of doped nickelates are sensitive to the small changes in $J/\Delta$. 
Interestingly indeed, recent full-band {\it ab initio} study reports higher weight of $|2,1\rangle$ than $|2,0\rangle$ \cite{BKang} indicating the effective enhancement of $J/\Delta$. 
Further studies are highly desirable to confirm our picture.

To conclude, we have identified distinctive correlated metal regimes emerging out of two-orbital Hubbard model at quarter-filling. This simple model is revealed to be differentiated from three- as well as five-orbital models in their Hund's metal active fillings, showing the transition to the VS phases. We found a generic scaling relation between the degree of spin-orbital separation ($\Delta T^\mathrm{onset}$) and $U_\mathrm{eff}$, and established a weak Hund's metal behavior in which $J$ enhances the correlation strength. 
We also discussed the implications of our two-orbital picture for the nature of metallic state of RE$_{1-\delta}$Sr$_\delta$NiO$_2$. We also remark that in this line of multiorbital viewpoint on nickelates, the role of nonlocal correlations/interactions and the emergence of symmetry-broken phases \cite{Arribi,Medici_2017,Steiner,Hoshino,Hoshino2,Ryee,Dumitrescu,Rodriguez} are intriguing open problems.
In addition to RE$_{1-\delta}$Sr$_\delta$NiO$_2$, the low-energy physics of {\it R}NiO$_3$ ({\it R}: rare-earth elements) is reported to be well described by Ni-$e_g$ bands \cite{Subedi,Seth}. Thus, a series of analysis presented in this study should also provide useful insights to these compounds.

{\it Acknowledgement}. S.R. and M.J.H. were supported by Creative Materials Discovery Program through NRF (2018M3D1A1058754) and Basic Science Research Program (2018R1A2B2005204). S.C. was supported by the U.S. Department of Energy, Office of Science, Basic Energy Sciences as a part of the Computational Materials Science Program. This research used resources of the National Energy Research Scientific Computing Center (NERSC), a U.S. Department of Energy Office of Science User Facility operated under Contract No. DE-AC02-05CH11231.


\bibliographystyle{apsrev}
\bibliography{ref}

\begin{thebibliography}{95}
\bibitem[{com({\natexlab{a}})}]{comment1}
\bibinfo{note}{For the region of VS insulators, we found strong oscilations
  between $N=2$ and $N=4$ solutions at lower temperatures during DMFT
  iterations, and thus stable convergence was not achieved. However, whole
  discussion of the present paper is not affected by this limitation.}
\bibitem[{com({\natexlab{b}})}]{comment2}
\bibinfo{note}{$Z= \big(1-{\partial \mathrm{Im}\Sigma(i\omega_n)}/{\partial
  \omega_n} \big|_{\omega_n \to +0}\big)^{-1}$, where $\Sigma(i\omega_n)$ is
  the fermionic self-energy on the Matsubara frequency axis. To evaluate $Z$,
  we fitted the fourth order polynomial to the lowest six Matsubara frequency
  $\Sigma(i\omega_n)$ following the strategy of Ref.~\cite{Mravlje}.}
\end{thebibliography}

\clearpage
\onecolumngrid

\setcounter{section}{0}
\setcounter{equation}{0}
\setcounter{figure}{0}

\renewcommand{\thesection}{S\arabic{section}}   
\renewcommand{\thetable}{S\arabic{table}}   
\renewcommand{\thefigure}{S\arabic{figure}}

\begin{center}
	\subsection*{\large
		Supplemental material for \\ ``Hund physics landscape of two-orbital system''
	}
	{Siheon Ryee,$^{1}$ Myung Joon Han,$^1$ and Sangkook Choi$^2$}\\
	\vspace{0.05in}
	\emph{$^{1}$Department of Physics, KAIST, Daejeon 34141, Republic of Korea}\\
	\emph{$^{2}$Condensed Matter Physics and Materials Science Department,\\ Brookhaven National Laboratory, Upton, NY 11973, USA}
\end{center}

\twocolumngrid
\vspace{0.1in}

\subsection*{Slater parametrization for five-orbital models}
We first consider on-site Coulomb interaction tensor defined by:
\begin{align}
\begin{split}
&\mathcal{U}_{m_1m_2m_3m_4} \\
&\equiv \int d\mathbf{r}d\mathbf{r}' \phi^*_{m_1}(\mathbf{r})\phi_{m_3}(\mathbf{r})V(\mathbf{r},\mathbf{r}') \phi^*_{m_2}(\mathbf{r}')\phi_{m_4}(\mathbf{r}'),
\end{split}
\end{align}
where $\phi_{m}(\mathbf{r})$ is a localized atomiclike basis function for orbital $m$. $V(\mathbf{r},\mathbf{r}')$ is a Coulomb potential which is assumed to be static for the present case. One widely adopted strategy to generate tensor elements is resorting to the following formula assuming rotational symmetry:
\begin{align} \label{S2}
\begin{split}
&\mathcal{U}_{m_1m_2m_3m_4} \\
&= \sum_{\{m_i'\}}\Big[S_{m_1m_1'}S_{m_2m_2'} \Big\{\sum_{k=0}^{2l}\alpha_k(m_1',m_2',m_3',m_4')F^k\Big\} \\ &\quad \times S^{-1}_{m_3'm_3}S^{-1}_{m_4'm_4} \Big].
\end{split}
\end{align}
Here $\alpha_k$ refers to Racah-Wigner numbers, $F^k$ to Slater integrals, and $l$ to angular momentum quantum number ($l=2$ for $d$-orbitals). $S$ is a transformation matrix from spherical harmonics to the predefined local basis sets. For the evaluation of $F^k$ in a solid environment, one requires advanced techniques, and thus it is often treated as a controllable parameter. 

As a demonstration, we list below matrix elements of intra/inter-orbital Coulomb interaction $\mathcal{U}_{mm'mm'}$ using cubic harmonics basis. Matrix elements are presented in the following ordering: $d_{x^2-y^2},d_{z^2},d_{xy},d_{yz},d_{xz}$ \cite{Pavarini}.
\begin{align} \label{S3}
\begin{split}
&\mathcal{U}_{mm'mm'}  \equiv \mathcal{U}_{mm'} =   \\
&\begin{pmatrix} \mathcal{U} & \mathcal{U}-2J_2 & \mathcal{U}-2J_3 & \mathcal{U} - 2J_1 & \mathcal{U} -2J_1  
\\ \mathcal{U}-2J_2 & \mathcal{U} & \mathcal{U}-2J_2 & \mathcal{U}-2J_4 & \mathcal{U}-2J_4 
\\ \mathcal{U}-2J_3 & \mathcal{U}-2J_2 & \mathcal{U} & \mathcal{U}-2J_1 & \mathcal{U}-2J_1
\\ \mathcal{U}-2J_1 & \mathcal{U}-2J_4 & \mathcal{U}-2J_1 & \mathcal{U} & \mathcal{U}-2J_1
\\ \mathcal{U}-2J_1 & \mathcal{U}-2J_4 & \mathcal{U}-2J_1 & \mathcal{U}-2J_1 & \mathcal{U}
\end{pmatrix},
\end{split}
\end{align}
where 
\begin{align}
\mathcal{U} &= F^0 + \frac{4}{49}(F^2+F^4) = F^0 + \frac{8}{5}\mathcal{J} \\
\mathcal{J} &= \frac{5}{98}(F^2+F^4) \\
J_1 &= \frac{3}{49}F^2 + \frac{20}{441}F^4 \\
J_2 &= -2\mathcal{J}+3J_1 \\
J_3 &= 6\mathcal{J}-5J_1 \\
J_4 &= 4\mathcal{J}-3J_1.
\end{align}
Matrix elements of exchange terms $\mathcal{U}_{mm'm'm}$ can also be evaluated from Eq.~(\ref{S2}) in a similar manner:
\begin{align} \label{S10}
\mathcal{U}_{mm'm'm}  \equiv \mathcal{J}_{mm'} =   \begin{pmatrix} \mathcal{U} & J_2 & J_3 & J_1 & J_1  
\\ J_2 &  \mathcal{U} & J_2 & J_4 & J_4 
\\ J_3 & J_2 & \mathcal{U} & J_1 & J_1
\\ J_1 & J_4 & J_1 & \mathcal{U} & J_1
\\ J_1 & J_4 & J_1 & J_1 & \mathcal{U}
\end{pmatrix}.
\end{align}
Note that when $\mathcal{J}=J_1$, or equivalently when $F^4/F^2=1.8$, elements of Eq.~(\ref{S3}) and Eq.~(\ref{S10}) become $\mathcal{U}$ when $m=m'$; $\mathcal{U}-2\mathcal{J}$ in Eq.~(\ref{S3}) and $\mathcal{J}$ in Eq.~(\ref{S10}) when $m\neq m'$, which leads to the form of Eq.~(1) in the main text. Hence, we can identify 
Kanamori parameters $U$ and $J$ as $U=\mathcal{U}=F^0+4(F^2+F^4)/49$ and $J=\mathcal{J}=3F^2/49+20F^4/441$, respectively, for five-orbital models. For most of $d$-orbital systems, however, $F^4/F^2$ is far from this Kanamori limit of $F^4/F^2=1.8$, approximately being $F^4/F^2\simeq 0.65$ for 3$d$ systems \cite{Vaugier}. 

In the main text, we followed the strategy of Strand \cite{Strand}, parametrizing $F^k$ as follows:
\begin{align}
F^0 &= U - \frac{8}{5}J \\
F^2 &= 49\big(\frac{1}{\gamma} + \frac{1}{7} \big)J \\
F^4 &= \frac{63}{5}J,
\end{align}
where $1/\gamma$ controls the relative strength of anisotropy of Coulomb interaction. In this way, $1/\gamma=0$ naturally corresponds to $F^4/F^2=1.8$ which is the Kanamori limit resulting in $U=\mathcal{U}$ and $J=\mathcal{J}$. For more realistic cases of $F^4/F^2\simeq 0.65$, $1/\gamma \simeq 1/4$. We finally remark that one should not be confused with the current definition of Hubbard $U$ and Hund's coupling $J$ with what is commonly adopted in most of literatures for five-orbital systems, which is $U=F^0$ and $J=(F^2+F^4)/14$.

\subsection*{$J$ vs. $U$ phase diagram and atomic multiplet probability.}

\begin{figure} [!htbp]
	\includegraphics[width=0.9\columnwidth, angle=0]{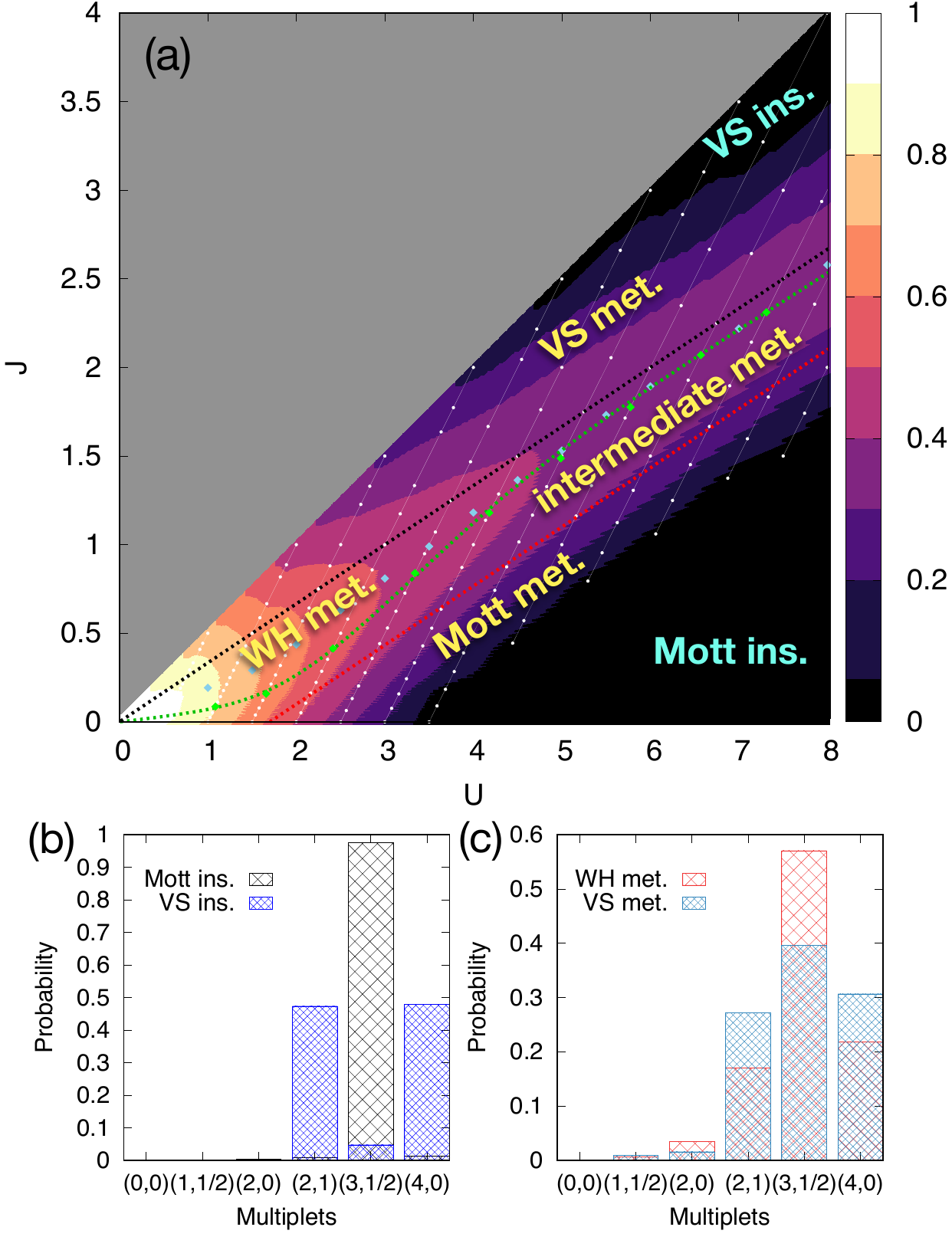}
	\caption{(a) $J$ vs. $U$ phase diagram of $Z$. 
		(b--c) Local multiplet probability profiles of four different regimes. Each multiplet is labeled in terms of $(N,S)$. $U$ and $J/U$ corresponding to each profile are $U=5,J/U=0.1$ ($U_\mathrm{eff}=3.5$) for Mott insulator, $U=5,J/U=0.5$ ($U_\mathrm{eff}=-2.5$) for VS insulator, $U=2,J/U=0.25$ ($U_\mathrm{eff}=0.5$) for WH metal, and $U=2,J/U=0.45$ ($U_\mathrm{eff}=-0.7$) for VS metal.}
	\label{sfig1}
\end{figure}

Figure~\ref{sfig1}(a) presents corresponding $J$ vs. $U$ phase diagram of Fig.~(3)(b) in the main text. One can notice the presence of two correlated insulators and metals intervening between the two insulators. Note that $dU_c(J)/dJ>0$ is clearly seen in this plot: a small increment of $J$ always pushes $U_c$ to a larger value.

In Fig.~\ref{sfig1}(b--c), we present multiplet probabilities. 
The Mott insulator is the result of suppressed charge fluctuations localizing electron motions. Thus $|N=3,S=1/2\rangle$ multiplet probability is maximized ($N$: charge, $S$: spin). By contrast, the VS insulator exhibits the predominance of two multiplets: $|2,1\rangle$ and $|4,0\rangle$ with largely suppressed $|3,1/2\rangle$ probability. 
In both WH and VS metals, the high spin $S=1$ is dominant in the $N=2$ charge subspace due to the effect of sizeable $J$ blocking the low-spin $S=0$ state. As VS metal emerges when $U_\mathrm{eff}<0$, multiplets in $|2,1\rangle$ and $|4,0\rangle$ are lower in energy than $|3,1/2\rangle$ exhibiting enhanced probabilitiy of $|2,1\rangle$ and $|4,0\rangle$ compared to the WH metal case.

We found that $T^\mathrm{onset}_\mathrm{spin}$ is distinctively higher in $U/D=1$ than the larger $U/D$ regions ($U/D\geq2$) as shown in Fig.~\ref{sfig1} exhibiting clear deviation of $U/D=1$ case from the rest. Considering that $T^\mathrm{onset}_\mathrm{spin}$ is a proxy for correlation strength, $U/D=1$ case should basically fall into the weakly correlated regime. Indeed large quasiparticle weight $Z$ is obtained near $U/D=1$ (see Fig.~\ref{sfig1}(a)).

\begin{figure}  [!htbp]
	\includegraphics[width=0.7\columnwidth, angle=0]{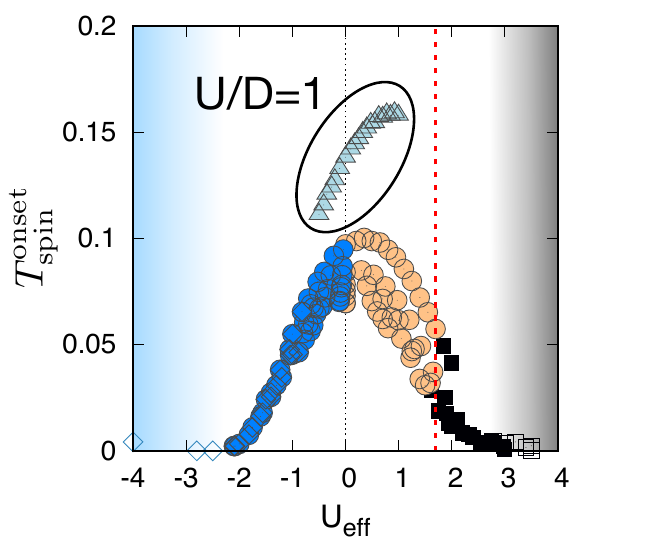}
	\caption{$T^\mathrm{onset}_\mathrm{spin}$ as a function of $U_\mathrm{eff}$. Filled triangles correspond to the $U/D=1$ case.}
	\label{sfig2}
\end{figure}


\subsection*{The effect other parametrizations of the local interaction I.}

\begin{figure} [!htbp]
	\includegraphics[width=0.7\columnwidth, angle=0]{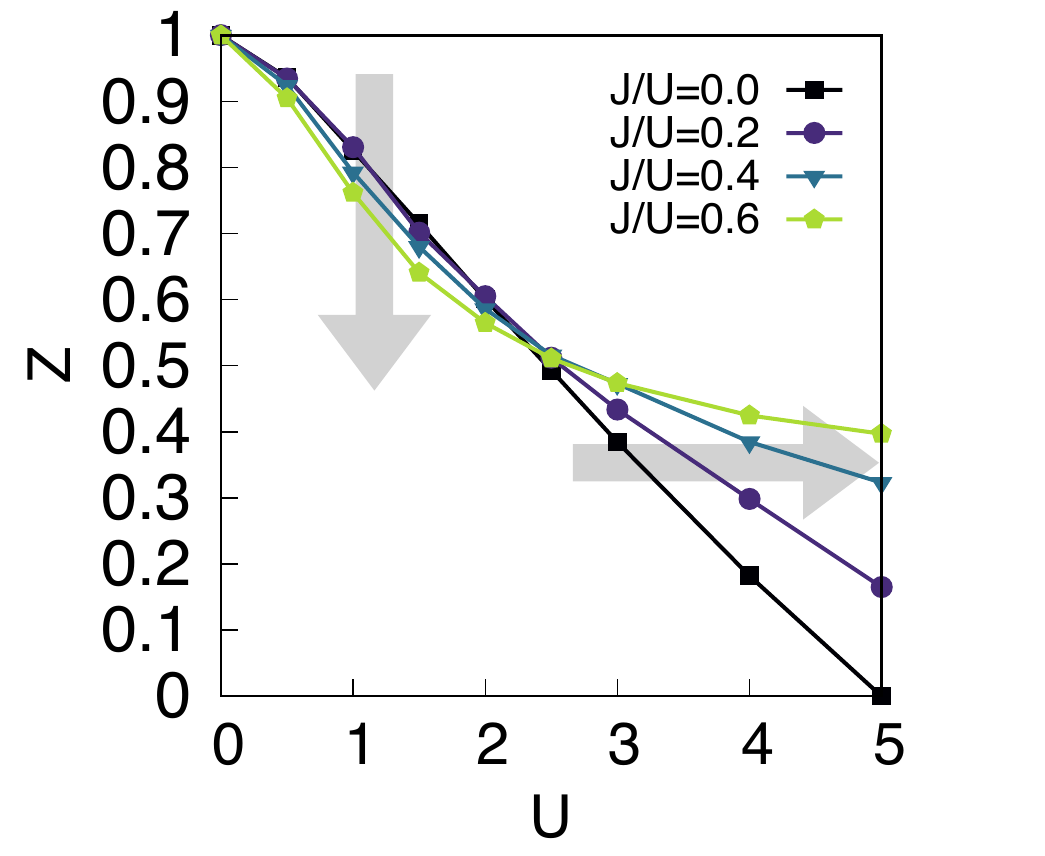}
	\caption{$Z$ as a function of $U$ with varying $J/U$. $U'=0.7U$.}
	\label{sfig3}
\end{figure}

In searching for the Janus effect, we argued that due to the choice of $U'=U-2J$, genuine effect of $J$ is masked by $U'$ which not fixed when $J$ is tuned. 
To test this idea, we set $U'=\alpha U$ with $\alpha$ being a tunable constant in Eq.~(1) in the main text. In this way, $U'$ as well as $U$ is disentangled from $J$, thereby more clear identification of their influences becomes feasible. As a demonstration, we present $Z$ obtained from $\alpha=0.7$ in Fig.~\ref{sfig3}. 
Note that in this case, $U_\mathrm{eff}=0.7U-J$. Thus $J/U\leq 0.7$ guarantees that $U_\mathrm{eff}\geq 0$. The Janus effect is observed, i.e, $(\partial{Z}/\partial{J})_{U}<0$ and $d{U_c}(J)/d{J}>0$, corroborating our conclusion that $J$ does induce this effect in two-orbital models.

\subsection*{The effect other parametrizations of the local interaction II.}

\begin{figure} [!htbp] 
	\includegraphics[width=0.99\columnwidth, angle=0]{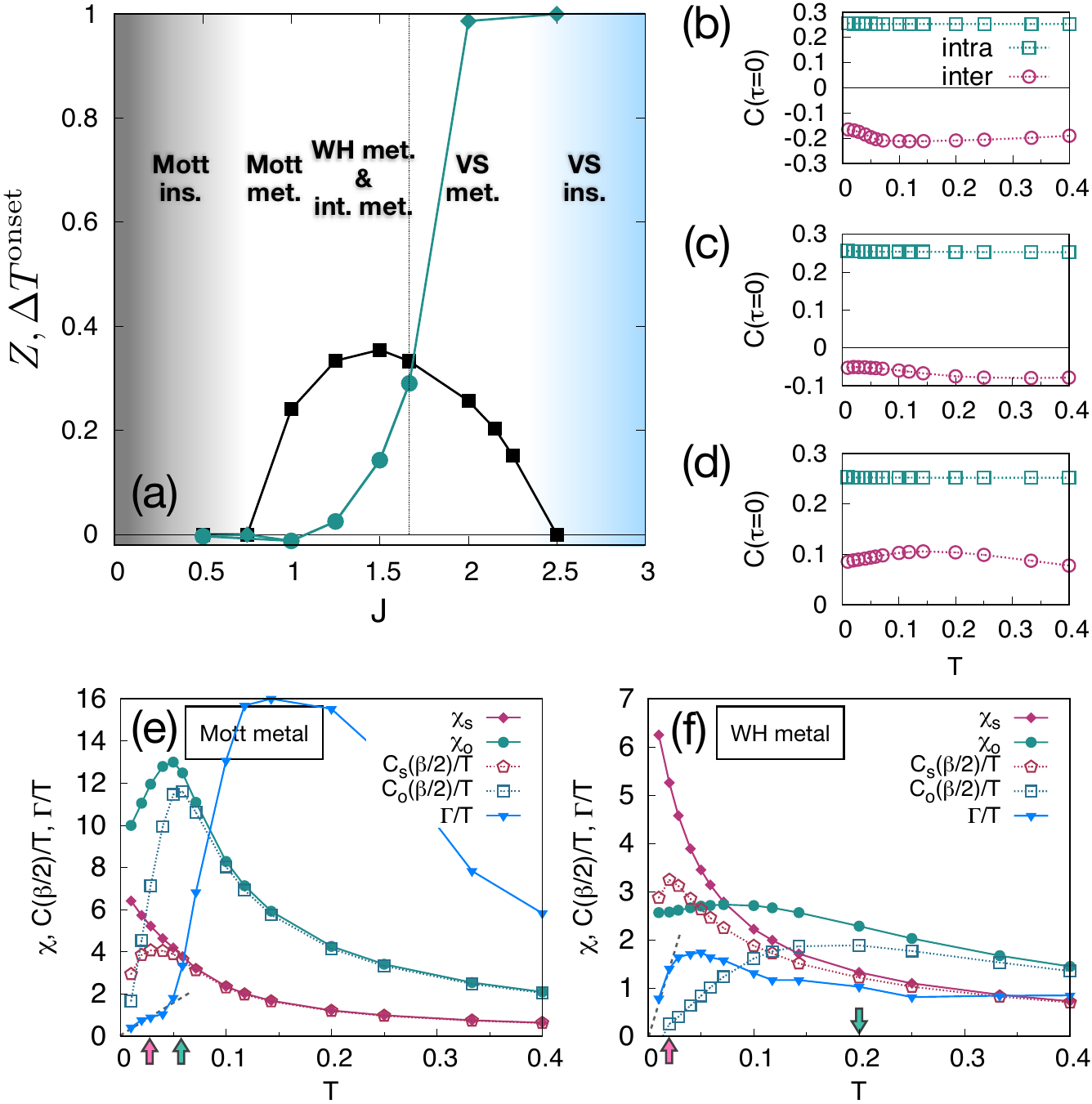}
	\caption{(a) $Z$ (black squares) and $\Delta T^\mathrm{onset}$ (green circles and diamonds) as a function of $J$ at $U=5$ using local interaction form of Eq.~(\ref{GK}) with $U'=U-2J$ and $J/2=J_\mathrm{X}=J_\mathrm{P}$.  The diamond symbol denotes the region in which $T^\mathrm{onset}_\mathrm{orb}>1$.
		(b--d) $C_\mathrm{intra}$ (green squares) and $C_\mathrm{inter}$ (magenta circles) at $U=5$ for (b) Mott metal ($J/U=0.2$; $U_\mathrm{eff}=2$), (c) WH metal ($J/U=0.3$; $U_\mathrm{eff}=0.5$), and (d) VS metal ($J/U=0.4$; $U_\mathrm{eff}=-1$).
		(e--f) $\chi_\mathrm{s/o}$, $\Gamma/T$, and $C_\mathrm{s/o}(\beta/2)/T$ plotted as a function of $T$ at $U=5$ for (e) Mott metal ($J/U=0.2$; $U_\mathrm{eff}=2$) and (f) WH metal ($J/U=0.3$; $U_\mathrm{eff}=0.5$). $T^\mathrm{peak}_\mathrm{spin/orb}$ are marked with magenta (spin) and green (orbital) arrows. The grey dashed lines are guide to the eye to indicate quasilinearity of $\Gamma/T$ (i.e., $\Gamma \sim T^2$). }
	\label{sfig4}
\end{figure}

The generalization of the local interaction Hamiltonian in Eq.~(1) in the main text reads
\begin{align}
\begin{split}
H_\mathrm{loc} &=  U\sum_{m}{n_{m \uparrow} n_{m \downarrow}} 
+ (U'-J\delta_{\sigma\sigma'})\sum_{m,m',\sigma,\sigma'}^{m < m'}{ n_{ m \sigma} n_{ m' \sigma'}} \\
&+ \sum_{m,m'}^{m \neq m'}(J_\mathrm{X}d^{\dagger}_{m \uparrow}  d^{\dagger}_{m' \downarrow} d_{m \downarrow} d_{m' \uparrow} 
+J_\mathrm{P}d^{\dagger}_{ m \uparrow} d^{\dagger}_{ m \downarrow} d_{ m' \downarrow} d_{ m' \uparrow}).
\end{split}
\label{GK}
\end{align}
The last two terms are now decoupled from $J$ which is responsible for the energy gain by forming the parallel spins (i.e., Hund's first rule).
Throughout the manuscript, we have considered the common choice of setting $U'=U-2J$ and $J=J_\mathrm{X}=J_\mathrm{P}$. To check the robustness of the existence of various metallic regimes discussed in the main text with respect to the choice of parametrization, we now consider a different setting: $U'=U-2J$ and $J/2=J_\mathrm{X}=J_\mathrm{P}$. In this case, both orbital and spin rotational symmetries are broken.
In any case, we obtained qualitatively the same phase diagram (see Fig.~\ref{sfig4}): as $J$ is increased from $J=0$, system evolves from Mott to VS insulator with four different intervening metallic regimes. Transition from WH to VS metal is found also at $U_\mathrm{eff}=U-3J=0$, i.e., $J\simeq1.67$ when $U=5$ as presented in Fig.~\ref{sfig4}(a). 

\begin{table*} [!htbp]
	\renewcommand{\arraystretch}{1.5}
	\begin{tabular}{c  c  c  c  c  c }
		\hline \hline
		&\ \ Eigenstates &\ \ $N$ &\ \ $S$ &\ \ $S_z$ &\ \ Eigenvalues \\
		\hline \hline	
		$|1\rangle$ &\ \ $|0,0\rangle$ &\ \ 0  &\ \ 0 &\ \  0 &\ \  0  \\
		\hline
		$|2\rangle$ &\ \ $|0,\uparrow\rangle$&\ \ 1 &\ \ 1/2 & \ \ 1/2 &\ \  $-\Delta-\mu$ \\
		$|3\rangle$ &\ \ $|0,\downarrow\rangle$&\ \ 1 &\ \ 1/2 & \ \ -1/2 &\ \  $-\Delta-\mu$ \\
		$|4\rangle$ &\ \ $|\uparrow,0\rangle$&\ \ 1 &\ \ 1/2 & \ \ 1/2 &\ \  $-\mu$ \\
		$|5\rangle$ &\ \ $|\downarrow,0\rangle$&\ \ 1 &\ \ 1/2 & \ \ -1/2 &\ \  $-\mu$ \\		
		\hline
		$|6\rangle$ &\ \ $|\uparrow,\uparrow\rangle$ &\ \ 2 &\ \ 1 &\ \ 1 &\ \  $U-3J-\Delta-2\mu$ \\ 
		$|7\rangle$ &\ \ $\big(|\uparrow,\downarrow\rangle + |\downarrow,\uparrow\rangle\big)/\sqrt{2}$ &\ \ 2 &\ \ 1 &\ \ 0 &\ \  $U-3J-\Delta-2\mu$ \\ 
		$|8\rangle$ &\ \ $|\downarrow,\downarrow\rangle$ &\ \ 2 &\ \ 1 &\ \ -1 &\ \  $U-3J-\Delta-2\mu$ \\
		$|9\rangle$ &\ \ $\big(|\uparrow,\downarrow\rangle - |\downarrow,\uparrow\rangle\big)/\sqrt{2}$ &\ \ 2 &\ \ 0 &\ \ 0 &\ \  $U-J-\Delta-2\mu$ \\ 
		$|10\rangle$ &\ \ $\frac{\alpha+\beta}{\sqrt{(\alpha+\beta)^2+J^2}}|\uparrow\downarrow,0\rangle - \frac{J}{\sqrt{(\alpha+\beta)^2+J^2}} |0,\downarrow\uparrow\rangle$ &\ \ 2 &\ \ 0 &\ \ 0 &\ \  $U-\Delta-\sqrt{J^2+\Delta^2}-2\mu$ \\ 
		$|11\rangle$ &\ \ $\frac{\alpha-\beta}{\sqrt{(\alpha-\beta)^2+J^2}}|\uparrow\downarrow,0\rangle - \frac{J}{\sqrt{(\alpha-\beta)^2+J^2}} |0,\downarrow\uparrow\rangle$ &\ \ 2 &\ \ 0 &\ \ 0 &\ \  $U-\Delta+\sqrt{J^2+\Delta^2}-2\mu$ \\ 
		\hline
		$|12\rangle$ &\ \ $|\uparrow,\uparrow\downarrow\rangle$&\ \ 3 &\ \ 1/2 & \ \ 1/2 &\ \  $3U-5J-2\Delta-3\mu$ \\
		$|13\rangle$ &\ \ $|\downarrow,\uparrow\downarrow\rangle$&\ \ 3 &\ \ 1/2 & \ \ -1/2 &\ \  $3U-5J-2\Delta-3\mu$ \\
		$|14\rangle$ &\ \ $|\uparrow\downarrow,\uparrow\rangle$&\ \ 3 &\ \ 1/2 & \ \ 1/2 &\ \  $3U-5J-\Delta-3\mu$ \\
		$|15\rangle$ &\ \ $|\uparrow\downarrow,\downarrow\rangle$&\ \ 3 &\ \ 1/2 & \ \ -1/2 &\ \  $3U-5J-\Delta-3\mu$ \\		
		\hline	   
		$|16\rangle$ &\ \ $|\uparrow \downarrow, \uparrow \downarrow \rangle$ &\ \ 4 &\ \ 0 &\ \ 0 &\ \ $6U-10J-2\Delta-4\mu$ \\
		\hline \hline
		
	\end{tabular}
	\caption{$\alpha=-\Delta$ and $\beta=\sqrt{J^2+\Delta^2}$. $U'=U-2J$, $\epsilon_1=0$, and $\epsilon_2=-\Delta$ in Eq.~(\ref{eq14}). The first entry in a ket of an eigenstate is the state of orbital-1 while the second is of orbital-2. }
	\label{table_s1}
\end{table*}

\subsection*{The effect of on-site energy level splitting and implications for RE$_{1-\delta}$Sr$_\delta$NiO$_2$}

We now consider the following local Hamiltonian for two orbitals:
\begin{align}
\begin{split}
H_\mathrm{loc} &=  U\sum_{m}{n_{m \uparrow} n_{m \downarrow}} 
+ \sum_{mm',\sigma\sigma'}^{m < m'}(U'-J\delta_{\sigma\sigma'}){ n_{ m \sigma} n_{ m' \sigma'}} \\
&+ J\sum_{m,m'}^{m \neq m'}(d^{\dagger}_{m \uparrow}  d^{\dagger}_{m' \downarrow} d_{m \downarrow} d_{m' \uparrow} 
+d^{\dagger}_{ m \uparrow} d^{\dagger}_{ m \downarrow} d_{ m' \downarrow} d_{ m' \uparrow}) \\
& +\sum_{m,\sigma}(\epsilon_m - \mu)n_{m\sigma},
\label{eq14}
\end{split}
\end{align}
where $\epsilon_m$ is the on-site energy level of orbital $m$ ($m=1,2$). $U'=U-2J$. Eigenvalues and corresponding eigenstates of Eq.~(\ref{eq14}) with a finite energy level splitting between two orbitals ($\epsilon_1=0$ and $\epsilon_2=-\Delta$) are presented in Table~\ref{table_s1}. Note that the lowest-energy spin state in $N=2$ subspace is determined by the ratio $J/\Delta$; eigenvalues of local Hamiltonian are $E_{S=1} = U-\Delta-\sqrt{J^2+\Delta^2}-2\mu$ for singlet $|10\rangle$ and $E_{S=0}=U-3J-\Delta-2\mu$ for triplet $|6\rangle$, $|7\rangle$, and $|8\rangle$. Hence, the criterion for predominance of triplet (i.e., $E_{S=1}<E_{S=0}$) is $J/\Delta>\sqrt{2}/4 \simeq 0.354$.

\begin{figure} [!htbp] 
	\includegraphics[width=0.99\columnwidth, angle=0]{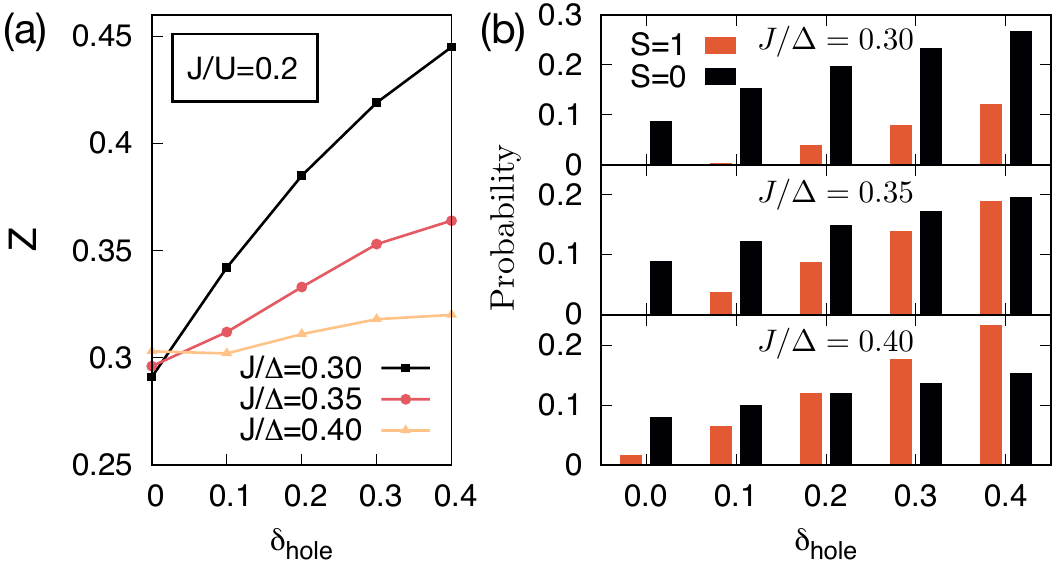}
	\caption{$\delta_\mathrm{hole}$ dependence of (a) $Z$ of orbital-1 at $U/D=2$ and $J/U=0.2$ with varying $\Delta$ and (b) corresponding atomic multiplet probabilities of $S=1$ (orange) and $S=0$ (black) states in the $N=2$ charge subspace.}
	\label{sfig5}
\end{figure}

With this insight, we further contemplate the effect of a finite $\Delta$ and the resulting spin state in the $N=2$ charge subspace (or equivalently, the two hole subspace) in connection with infinite-layer nickelates RE$_{1-\delta}$Sr$_\delta$NiO$_2$. We take $U/D \simeq 2$ and $J/U = 0.2$ relevant for a Ni-$e_g$ model of Nd$_{0.8}$Sr$_{0.2}$NiO$_2$ \cite{Sakakibara}. 
Figure~\ref{sfig5}(a--b) present the hole doping ($\delta_\mathrm{hole}$) dependence of $Z$ of orbital-1 and the atomic multiplet probabilities of triplet $S=1$ and singlet $S=0$ states in the $N=2$ charge subspace with varying $\Delta$ obtained from our DMFT calculations.

One can notice from Fig.~\ref{sfig5}(b) that being consistent with atomic multiplet analysis, $S=1$ starts to prevail over $S=0$ when $J/\Delta > 0.354$ as $\delta_\mathrm{hole}$ increases. However, singlet $S=0$ is predominant at the relevant value of $J/\Delta \simeq 0.3$ for Nd$_{0.8}$Sr$_{0.2}$NiO$_2$.

In any case, we found that $Z$ is enhanced as $\delta_\mathrm{hole}$ is introduced Fig.~\ref{sfig5}(a), albeit this enhancement is gradually diminished as $J/\Delta$ is increased. 
Interestingly, $\delta_\mathrm{hole}$ dependence of $Z$ obtained from our two-orbital model (Fig.~\ref{sfig5}(a)) is consistent with several previous {\it ab initio} results reporting the enhancement of $Z$ of Ni-$d_{x^2-y^2}$ state upon hole doping \cite{Kitatani,YWang,Petocchi,BKang}. Thus, it may be interpreted as a signature showing that the singlet rather than triplet dominates the two-hole subspace of RE$_{1-\delta}$Sr$_\delta$NiO$_2$. This supports Mott's metallic behavior, in line with an interpretation of recent x-ray absoption spectroscopy (XAS) data \cite{Rossi}.

However, there is another way to explain the hole-doping induced $Z$ enhancement: $\delta_\mathrm{hole}$-induced $\Delta$ enhancement. Within \textit{ab initio} linearized quasiparticle self-consistent GW~+~DMFT (LQSGW+DMFT) \cite{BKang}, $\Delta$ is enhanced by 20~$\%$ upon 0.2 hole doping. Also interestingly, \textit{ab initio} LQSGW+DMFT approach reports approximately two times larger probability of $S=1$ than that of $S=0$ in Ni-$e_g$ subspace \cite{BKang}. These results suggest that extrinsic doping changes model Hamiltonian parameters. Besides, it has been reported that the majority of the doped hole goes into other orbitals than Ni-$d$ \cite{BKang}. At $n_d=2.6$ ($\delta_\mathrm{hole}=0.4$)  which corresponds the Ni-$e_g$ occupation of RE$_{0.8}$Sr$_{0.2}$NiO$_2$ obtained from \textit{ab initio} LQSGW+DMFT, $Z$ increases as $\Delta$ increases as shown in Fig.~\ref{sfig5}(a). The Hund's metal picture, which is not compatible with the aforementioned doping dependence under fixed parameters, may be reconciled with the doping-induced $Z$ enhancement in this way.

\end{document}